\documentclass[runningheads]{llncs}
\usepackage{cite}
\usepackage{amsmath,amssymb,amsfonts}
\usepackage{algorithmic}
\usepackage{graphicx}
\usepackage{textcomp}
\usepackage{xcolor}
\usepackage{graphicx}
\usepackage{algorithm} 
\usepackage{algorithmic} 
\usepackage{graphicx}
\usepackage{float}
\usepackage{subfigure}
\begin{document}
\title{Neural Networks in Hybrid Precoding for Millimeter Wave Massive MIMO Systems\thanks{This work was supported by the National Key Research and Development Program of China under Grant 2017YFE0121600.}}
\titlerunning{Neural Networks in HP for mmWave Massive MIMO Systems}
%
\author{Jing Yang\inst{1}\orcidID{0000-0003-4639-6592} \and
Kai Chen\inst{1} \and
Xiaohu Ge\inst{1}\orcidID{0000-0002-3204-5241} \and
Yonghui Li\inst{2} \and
Lin Tian\inst{3}}
\authorrunning{J. Yang et al.}
%
\institute{School of Electronic Information and Communications, Huazhong University of Science and Technology, Wuhan, Hubei, China\\
\email{xhge@mail.hust.edu.cn}\\
\and
School of Electrical and Information Engineering, University of Sydney, Sydney, Australia\\
\and
Beijing Key Laboratory of Mobile Computing and Pervasive Devices, Institute of Computing Technology, Chinese Academy of Sciences, China\\
}
\maketitle              
\begin{abstract}
Neural networks have been applied to the physical layer of wireless communication systems to solve complex problems. In millimeter wave (mmWave) massive multiple-input multiple-output (MIMO) systems, hybrid precoding has been considered as an energy-efficient technology to replace fully-digital precoding. The way of designing hybrid precoding in mmWave massive MIMO systems by multi-layer neural networks has not been investigated. Based on further decomposing the baseband precoding matrix, an idea is proposed in this paper to map hybrid precoding structure to a multi-layer neural network. Considering the deterioration in the throughput and energy efficiency of mmWave massive MIMO systems, the feasibility of the proposed idea is analyzed. Moreover, a singular value decomposition (SVD) based decomposing (SVDDE) algorithm is proposed to evaluate the feasibility of the proposed idea. Simulation results indicate that there is an optimal number of users which can minimize the performance deterioration. Moreover, the simulation results also show that slight deterioration in the throughput and energy efficiency of mmWave massive MIMO systems is caused by further decomposing the baseband precoding matrix. In other words, further decomposing the baseband precoding matrix is a feasible way to map the hybrid precoding structure to a multi-layer neural network.

\end{abstract}
\section{Introduction}
The design and optimization of the fifth generation (5G) wireless communication system become challenging, due to the expectation of satisfying the key performance indicators (KPIs) in 5G usage scenarios, such as 100 Mbit/s user experienced data rate, 10 Gbit/s peak data rate, 1 millisecond (ms) over-the-air latency and ${{10}^{6}}/\text{k}{{\text{m}}^{2}}$ connectivity density \cite{key-1,key-2}. Combining the millimeter wave (mmWave) communication and massive multiple-input multiple-output (MIMO) technology is a feasible solution to meet the KPIs \cite{key-3}. Meanwhile, the signal processing of baseband units (BBUs) becomes more complicated in 5G base stations (BSs) \cite{key-4}. In this case, it is intractable to optimize the real-time hybrid precoding in multi-user mmWave massive MIMO systems \cite{key-5,key-6}. Neural networks, one of the technologies in artificial intelligence (AI), have shown the great application value to solve complex and intractable problems in image recognition, automatic control and healthcare \cite{key-7,key-8, key-9}. Therefore, it is attractive to apply neural networks to design hybrid precoding for multi-user mmWave massive MIMO systems.

Some studies have already investigated the application of neural networks for the physical layer of wireless communication systems. The work in \cite{key-10} proposed a procedure to predict channel characteristics of mmWave massive MIMO systems, based on convolutional neural networks. Moreover, the predicted results in \cite{key-10} showed the well matching with the real channel characteristics. In the channel estimation of mmWave massive MIMO systems, the estimation with the help of neural networks was better than the state-of-the-art compressed sensing algorithms \cite{key-11}. In addition to the channel estimation, neural networks were also used for the modulation classification of raw IQ samples, which achieved competitive accuracy \cite{key-12}. Considering the unmanageable joint optimization problem of the coverage and capacity in mmWave massive MIMO systems, the authors in \cite{key-13} enhanced the service coverage and the spectrum efficiency by applying neural networks to solve the joint optimization problem. Combining the distributed massive MIMO with neural networks, more accurate results of user positioning had been achieved \cite{key-14}, which paves the way for network operators to provide better context-aware communication services. It is emerging in designing the physical layer of 5G wireless communication systems by neural networks. The work in \cite{key-15} provided a comprehensive survey of applications which uses multi-layer neural networks to solve problems in cellular networks. Based on the results in \cite{key-15}, the existing studies, related to the application of neural networks in the physical layer of cellular networks, is divided into five categories, i.e., signal detection, modulation classification, error correction, interference alignment management and anti-jamming. Applications of neural networks in wireless communication systems remain to be explored.

Precoding is one of the key technologies in the physical layer of massive MIMO systems to improve the spectrum efficiency. Considering the high cost of radio frequency (RF) chains in mmWave band for fully-connected precoding, hybrid precoding is proposed in mmWave massive MIMO systems \cite{key-16}. The topology of the fully-connected phase shifter network in hybrid precoding structure is similar to neural networks. To our knowledge, studies related to design hybrid precoding by multi-layer neural networks have not appeared in the available literature. Inspired by this vacancy in knowledge, the objective of this paper is proposing an idea to map the hybrid precoding structure to multi-layer neural networks. The idea is based on the further decomposition of the baseband precoding matrix in hybrid precoding. Furthermore, an SVD-based decomposing (SVDDE) algorithm is proposed to evaluate the feasibility of the proposed idea. The simulation results show that the performance deterioration in the throughput and energy efficiency of mmWave massive MIMO systems can be caused when the baseband precoding matrix in hybrid precoding is further decomposed. In this case, modeling the hybrid precoding structure as multi-layer neural networks is feasible in mmWave massive MIMO systems.

The rest of this paper is outlined as follows. Section II describes the system model. In Section III, an idea is proposed to map hybrid precoding to a multi-layer neural network. Moreover, the feasibility of the proposed idea is analyzed. Section IV provides the simulation results. Finally, conclusions are drawn in Section V.

\section{System Model}

\begin{figure}[htbp]
\centerline{\includegraphics[width=\textwidth]{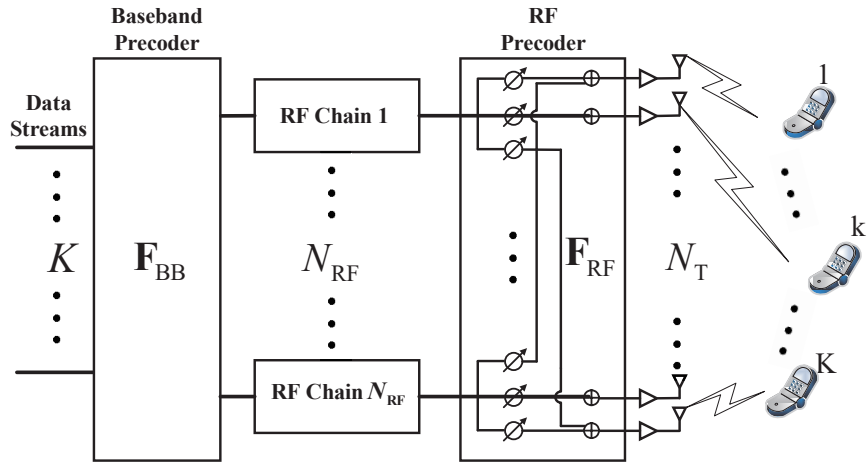}}
\caption{The mmWave massive MIMO systme with fully-connected hybrid precoding structure.}
\label{Fig1}
\end{figure}

As shown in Fig. \ref{Fig1}, $K$ single antenna users are served by the mmWave massive MIMO system with fully-connected hybrid precoding. The BS is equipped with ${{N}_{\text{T}}}$ antennas and ${{N}_{\text{RF}}}$ RF chains. Moreover, the values of $K$, ${{N}_{\text{T}}}$ and ${{N}_{\text{RF}}}$ satisfy $K\le {{N}_{\text{RF}}}\le {{N}_{\text{T}}}$. The phase shifter network in Fig. \ref{Fig1} contains ${{N}_{\text{T}}}{{N}_{\text{RF}}}$ phase shifters (PSs). The baseband precoding matrix is denoted as ${{\mathbf{F}}_{\text{BB}}}\in {{\mathbb{C}}^{{{N}_{\text{RF}}}\times K}}$ and the RF precoding matrix is denoted as ${{\mathbf{F}}_{\text{RF}}}\in {{\mathbb{C}}^{{{N}_{\text{T}}}\times {{N}_{\text{RF}}}}}$. The downlink channel matrix $\mathbf{H}\in {{\mathbb{C}}^{{{N}_{\text{T}}}\times K}}$ is ${{\mathbf{H}}^{\text{H}}}={{\left[ {{\mathbf{h}}_{1}},\cdots ,{{\mathbf{h}}_{k}},\cdots ,{{\mathbf{h}}_{K}} \right]}^{\text{H}}}$, where ${{\mathbf{h}}_{k}}$ is the downlink channel vector between the BS and the $k\text{-th}$ user. The received signal at the $k\text{-th}$ user is given as
\begin{equation}
{{y}_{k}}=\mathbf{h}_{k}^{\text{H}}{{\mathbf{F}}_{\text{RF}}}{{\mathbf{F}}_{\text{BB}}}\mathbf{s}+{{n}_{k}},
\end{equation}
where $\mathbf{s}$ is $K\times 1$ transmitted signal vector $\mathbf{s}={{\left[ {{s}_{1}},\cdots ,{{s}_{k}},\cdots ,{{s}_{K}} \right]}^{\text{T}}}$ for $K$ users satisfying $\mathbb{E}\left[ \mathbf{s}{{\mathbf{s}}^{\text{H}}} \right]={{\mathbf{I}}_{K}}$, ${{s}_{k}}$ ($k=1,\cdots ,K$) is the transmitted signal for the $k\text{-th}$ user. ${{n}_{k}}\sim \mathcal{C}\mathcal{N}\left( 0,\sigma _{n}^{2} \right)$ is the noise received by the $k\text{-th}$ user. ${{\mathbf{F}}_{\text{BB}}}$ and ${{\mathbf{F}}_{\text{RF}}}$ satisfy $\left\| {{\mathbf{F}}_{\text{RF}}}{{\mathbf{F}}_{\text{BB}}} \right\|_{F}^{2}={{P}_{\text{T}}}$ where ${{P}_{\text{T}}}$ is the transmission power.

The Saleh-Valenzuela channel model is considered as the mmWave channel model \cite{key-17} and ${{\mathbf{h}}_{k}}$ is given as
\begin{equation}
{{\mathbf{h}}_{k}}=\sqrt{\frac{{{N}_{\text{T}}}{{\xi }_{k}}}{L}}\sum\limits_{l=1}^{L}{g_{l}^{k}\mathbf{a}\left( {{\theta }_{l}} \right)},
\end{equation}
where $L$ is the total number of multipath between the BS and $K$ users. The large-scale fading coefficient is denoted as ${{\xi }_{k}}=\frac{1}{d_{k}^{\alpha }}$, where $\alpha $ is the path loss exponent of mmWave and ${{d}_{k}}$ is the distance between the BS and the $k\text{-th}$ user \cite{key-18}. $g_{l}^{k}\sim \mathcal{C}\mathcal{N}\left( 0,\sigma _{g,\ l}^{2} \right)$ is the complex gain of signals at the $l\text{-th}$ multipath \cite{key-19}. The array response vector $\mathbf{a}\left( {{\theta }_{l}} \right)$ of the uniform linear array (ULA) in Fig. \ref{Fig1} is written as
\begin{equation}
\mathbf{a}\left( {{\theta }_{l}} \right)=\frac{1}{\sqrt{{{N}_{\text{T}}}}}{{\left[ 1,{{e}^{j\frac{2\pi }{\lambda }{{d}_{\text{T}}}\sin \left( {{\theta }_{l}} \right)}},\cdots ,{{e}^{j\left( {{N}_{\text{T}}}-1 \right)\frac{2\pi }{\lambda }{{d}_{\text{T}}}\sin \left( {{\theta }_{l}} \right)}} \right]}^{\text{T}}},
\end{equation}
where ${{\theta }_{l}}$ is the azimuth angle of signals at the $l\text{-th}$ multipath, $\lambda $ is the wavelength of mmWave, ${{d}_{\text{T}}}$ is the inter-antenna spacing.

\section{The Neural Network in Hybrid Precoding}
\subsection{The Mapping of A Multi-Layer Neural Network}
\begin{figure}[htbp]
\subfigure[]{
\label{a}
\includegraphics[width=\textwidth]{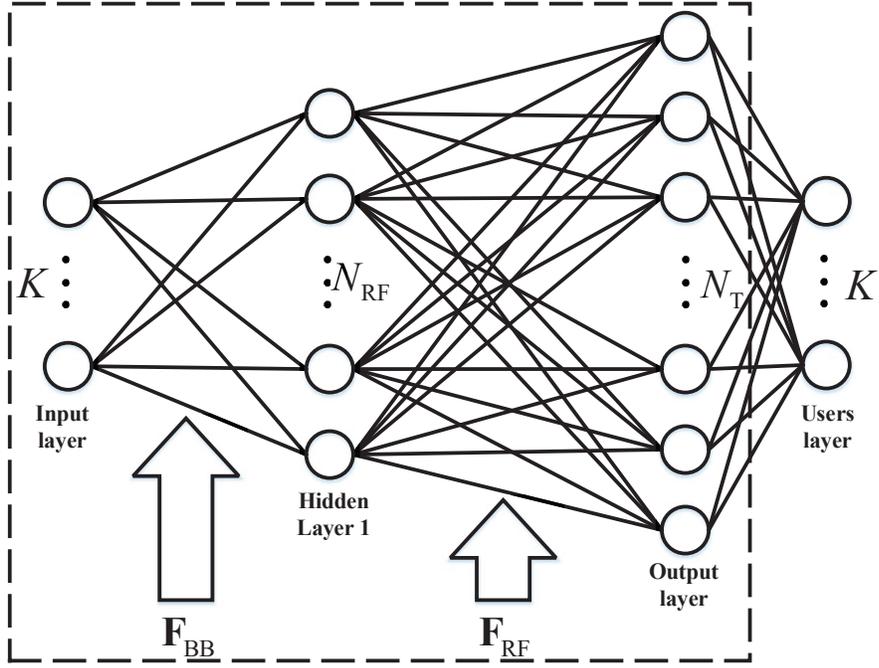}}
\subfigure[]{
\label{b}
\includegraphics[width=\textwidth]{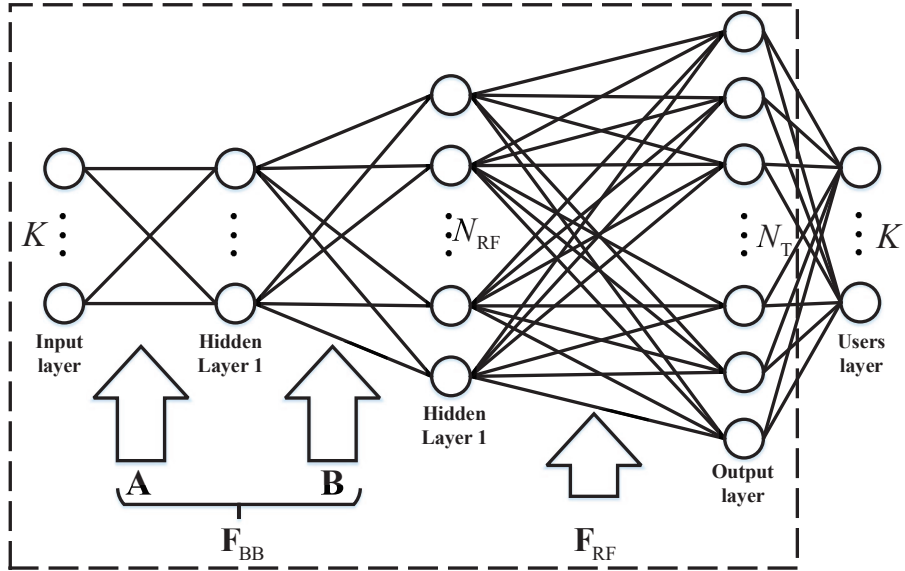}}
\caption{Hybrid precoding structure in mmWave massive MIMO system is mapped to (a) a one-layer neural network; (b) a multi-layer neural network.}\label{Fig2}
\end{figure}

Most of the studies related to neural networks are training software-based neural networks. Few works have been done in training hardware-based neural networks to achieve learning \cite{key-20}. In fully-connected hybrid precoding structure (Fig. \ref{Fig1}), the phase shifter network has similar topology to neural networks. Furthermore, the baseband precoding processing in BBUs is described as the matrix-vector multiplication ${{\mathbf{F}}_{\text{BB}}}\mathbf{s}$, which is similar to the mathematic model of neural networks, i.e., the multiplication between weight matrices and input data vectors. It is reasonable to map the hybrid precoding structure in Fig. \ref{Fig1} to a one-layer neural network in Fig. \ref{Fig2}a.

The essence of precoding is the signal processing between the input data streams and antennas array. The input data streams and antennas array are treated as the input layer and output layer, respectively, for the equivalent neural network in Fig. \ref{Fig2}. In Fig. \ref{Fig2}a, the hybrid precoding structure in mmWave massive MIMO systems is mapped to a single-hidden-layer neural network, whose weight matrices are ${{\mathbf{F}}_{\text{BB}}}$ and ${{\mathbf{F}}_{\text{RF}}}$. In hybrid precoding, the digital precoding can adjust both the amplitude and phase of signals and the RF precoding can only change the phase of signals, which implies that the mapped neural network in Fig. \ref{Fig2}a has two weight matrices with different properties. There is no denying that multi-layer neural networks commonly have better performance than one layer neural networks. It is reasonable to map the hybrid precoding structure to a multi-layer neural networks. More than two weight matrices are needed in mapping the hybrid precoding structure to a multi-layer neural network.  In addition to matrices ${{\mathbf{F}}_{\text{BB}}}$ and ${{\mathbf{F}}_{\text{RF}}}$, new weight matrices have to be obtained.

Hybrid precoding is proposed by decomposing fully-digital precoding into a digital baseband precoding and an analog RF precoding. Inspired by the decomposition of fully-digital precoding, an idea that mapping the hybrid precoding structure to a multi-layer neural network is proposed. The baseband precoding matrix ${{\mathbf{F}}_{\text{BB}}}$ is first decomposed into two new sub-matrices. These new sub-matrices are treated as new weight matrices for a two-layer neural network in Fig. \ref{Fig2}b. Different from software-based neural networks and hardware-based neural networks, the equivalent neural network in Fig. \ref{Fig2}b is a software-hardware hybrid neural network which consists of a software-based neural network in BBUs and a hardware-based neural network in the phase shifter network. Software-based neural networks have been widely implemented. In contrast, the implementation of hardware-based neural networks still has knotty technical problems. One formidable challenge for implementations of the hardware-based neural network in Fig. \ref{Fig2} is to achieve learning in the phase shifter network. Considering that the software-hardware hybrid neural network has not been implemented, we only investigate the feasibility of the proposed idea.

Considering that the proposed idea is based on further decomposing ${{\mathbf{F}}_{\text{BB}}}$, we speculate on the feasibility of the proposed idea by using existing methods, due to the difficulty of implementing the software-hardware hybrid neural network. The proposed idea is feasible at the case that the performance deterioration, caused by the decomposition of ${{\mathbf{F}}_{\text{BB}}}$, is slight in the throughput and energy efficiency of mmWave massive MIMO systems. Considering a widely used approach to decompose ${{\mathbf{F}}_{\text{BB}}}$, the decomposition algorithm should be proposed and the impact of further decomposing ${{\mathbf{F}}_{\text{BB}}}$ on the throughput and energy efficiency of mmWave massive MIMO systems with hybrid precoding structure has to be analyzed. The throughput and energy efficiency of mmWave massive MIMO systems is derived in the following.

The received signal-to-interference-plus-noise ratio (SINR) at the $k\text{-th}$ user is calculated as
\begin{equation}
SIN{{R}_{k}}=\frac{{{\left| \mathbf{h}_{k}^{\text{H}}{{\mathbf{F}}_{\text{RF}}}{{\mathbf{F}}_{\text{BB,}\ k}}\mathbf{F}_{\text{BB,}\ k}^{\text{H}}\mathbf{F}_{\text{RF}}^{\text{H}}{{\mathbf{h}}_{k}} \right|}^{2}}}{\sigma _{n}^{2}+\sum\limits_{i=1,i\ne k}^{K}{{{\left| \mathbf{h}_{k}^{\text{H}}{{\mathbf{F}}_{\text{RF}}}{{\mathbf{F}}_{\text{BB,}\ i}}\mathbf{F}_{\text{BB,}\ i}^{\text{H}}\mathbf{F}_{\text{RF}}^{\text{H}}{{\mathbf{h}}_{k}} \right|}^{2}}}},
\end{equation}
where ${{\mathbf{F}}_{\text{BB,}\ k}}$ is the $k\text{-th}$ column vector of ${{\mathbf{F}}_{\text{BB}}}$. Thus, the throughput for the BS to simultaneously serve $K$ users is presented by
\begin{equation}
{{R}_{\text{sum}}}=W\sum\limits_{k=1}^{K}{{{\log }_{2}}\left( 1+SIN{{R}_{k}} \right)},
\end{equation}
where $W$ is the bandwidth.

In Fig. \ref{Fig1}, the mmWave massive MIMO system with hybrid precoding structure consists of power amplifiers (PAs), the phase shifter network, RF chains and BBUs. The total power consumption of the mmWave massive MIMO system is
\begin{equation}
{{P}_{\text{total}}}={{P}_{\text{PA}}}+{{N}_{\text{T}}}{{N}_{\text{RF}}}{{P}_{\text{PS}}}+{{N}_{\text{RF}}}{{P}_{\text{RF}}}+{{P}_{\text{BB}}},
\end{equation}
where ${{P}_{\text{PA}}}=\frac{{{P}_{\text{T}}}}{{{\eta }_{\text{PA}}}}$ is the power of PAs, ${{\eta }_{\text{PA}}}$ is the efficiency of PAs \cite{key-21, key-22}. The power consumption of the phase shifter network is ${{N}_{\text{T}}}{{N}_{\text{RF}}}{{P}_{\text{PS}}}$, where ${{P}_{\text{PS}}}$ is the power of a single PS. The power consumption of RF chains is ${{N}_{\text{RF}}}{{P}_{\text{RF}}}$, where ${{P}_{\text{RF}}}$ is the power of one RF chain. The power consumed by BBUs can be obtained by
\begin{equation}
{{P}_{\text{BB}}}=W\frac{\Delta }{{{L}_{\text{BS}}}}+\frac{W}{{{W}_{\text{c}}}{{T}_{\text{c}}}}\frac{\Omega }{{{L}_{\text{BS}}}},
\end{equation}
where the first term is the power consumption of the multiplication ${{\mathbf{F}}_{\text{BB}}}\mathbf{s}$ in precoding processing, $\Delta $ is the number of floating-point operations for the multiplication ${{\mathbf{F}}_{\text{BB}}}\mathbf{s}$. The second term in (7) is the power consumption of the precoding algorithm used in BBUs, $\Omega $ is the number of floating-point operations for the precoding algorithm. The coherence bandwidth and coherence time of mmWave frequency are denoted as ${{W}_{\text{c}}}$ and ${{T}_{\text{c}}}$, respectively. The typical values of ${{W}_{\text{c}}}$ and ${{T}_{\text{c}}}$ in mmWave massive MIMO systems are $100\ \text{MHz}$ and $35\ \mu \text{s}$, respectively \cite{key-23, key-24}. 	${{L}_{\text{BS}}}$ is defined as the computation efficiency of BBUs, whose typical value is $12.8\ \text{GFLOPS}/\text{W}$ \cite{key-25}.

Based on (5) and (6), the energy efficiency of the mmWave massive MIMO system with hybrid precoding structure is denoted as
\begin{equation}
{{\eta }_{\text{EE}}}=\frac{{{R}_{\text{sum}}}}{{{P}_{\text{total}}}}.
\end{equation}

\subsection{The Decomposition Based on SVD}
Assuming that the BS has the perfect channel state information (CSI), the near-optimal baseband precoding matrix for maximizing the ${{R}_{\text{sum}}}$ is equivalent zero-forcing (ZF) precoding \cite{key-5} which is given by
\begin{equation}
\mathbf{F}_{\text{BB}}^{\text{opt}}=\mathbf{H}_{\text{eq}}^{\text{H}}{{\left( {{\mathbf{H}}_{\text{eq}}}\mathbf{H}_{\text{eq}}^{\text{H}} \right)}^{-1}}\mathbf{D},
\end{equation}
where ${{\mathbf{H}}_{\text{eq}}}={{\mathbf{H}}^{\text{H}}}\mathbf{F}_{\text{RF}}^{\text{opt}}$ is the $K\times {{N}_{\text{RF}}}$ equivalent downlink channel matrix for $K$ users, $\mathbf{F}_{\text{RF}}^{\text{opt}}$ is the optimal RF precoding matrix which can be written as linear combination of $\mathbf{a}\left( {{\theta }_{l}} \right)$, e.g., vectors $\mathbf{a}\left( {{\theta }_{l}} \right)$ with different ${{\theta }_{l}}$ as columns of $\mathbf{F}_{\text{RF}}^{\text{opt}}$ \cite{key-16}. $\mathbf{D}$ is a $K\times K$ diagonal matrix to normalize $\mathbf{F}_{\text{BB}}^{\text{opt}}$.
The singular value decomposition (SVD) is widely used to decompose a matrix. Accordingly, the SVD of $\mathbf{F}_{\text{BB}}^{\text{opt}}$ is considered as the decomposition approach to investigate the feasibility of the proposed idea. Define the SVD of $\mathbf{F}_{\text{BB}}^{\text{opt}}$ as $\mathbf{F}_{\text{BB}}^{\text{opt}}=\mathbf{U\Sigma }{{\mathbf{V}}^{\text{H}}}$, where $\mathbf{U}=\left[ {{\mathbf{u}}_{1}},{{\mathbf{u}}_{2}},\cdots ,{{\mathbf{u}}_{K}} \right]$ is a ${{N}_{\text{RF}}}\times K$ left singular vector matrix and $\mathbf{V}=\left[ {{\mathbf{v}}_{1}},{{\mathbf{v}}_{2}},\cdots ,{{\mathbf{v}}_{K}} \right]$ is a $K\times K$ right singular vector matrix. $\mathbf{\Sigma }=\text{diag}\left( {{\sigma }_{1}},{{\sigma }_{2}},\cdots ,{{\sigma }_{K}} \right)$ is a $K\times K$ diagonal matrix containing the nonzero singular values of $\mathbf{F}_{\text{BB}}^{\text{opt}}$ in decreasing order, i.e., ${{\sigma }_{1}}\ge {{\sigma }_{2}}\ge \cdots \ge {{\sigma }_{K}}$. Set $\mathbf{A}=\mathbf{U\Sigma }$ and $\mathbf{B}={{\mathbf{V}}^{\text{H}}}$, $\mathbf{F}_{\text{BB}}^{\text{opt}}$ is rewritten as $\mathbf{F}_{\text{BB}}^{\text{opt}}=\mathbf{AB}$. Decomposing $\mathbf{F}_{\text{BB}}^{\text{opt}}$ into two sub-matrices $\mathbf{A}$ and $\mathbf{B}$, a new hidden layer is added to the equivalent neural network which is a two-layer neural network in Fig. \ref{Fig2}b. Moreover, both matrices $\mathbf{A}\in {{\mathbb{C}}^{{{N}_{\text{RF}}}\times K}}$ and $\mathbf{B}\in {{\mathbb{C}}^{K\times K}}$ are treated as the weight matrices for the two-layer neural network.

Combining the mmWave communication and massive MIMO technologies in 5G BSs complicates the real-time signal processing in BBUs. Moreover, the number of floating-point operations in precoding processing has been increased, which increases the power consumption of BBUs \cite{key-4}. Considering that the power consumption of BBUs has a great impact on the energy efficiency of mmWave massive MIMO systems, the decomposition of $\mathbf{F}_{\text{BB}}^{\text{opt}}$ should not increase the power consumption of BBUs, i.e., not increasing the number of floating-point operations in hybrid precoding. The number of floating-point operations for multiplications ${{\mathbf{F}}_{\text{BB}}}\mathbf{s}$ and $\mathbf{ABs}$ are ${{\Lambda }_{1}}=8{{N}_{\text{RF}}}K-2{{N}_{\text{RF}}}$ and ${{\Lambda }_{2}}=8\left( {{N}_{\text{RF}}}K+{{K}^{2}} \right)-2\left( {{N}_{\text{RF}}}+K \right)$, respectively. The additional number of floating-point operations in hybrid precoding processing is denoted as $\Phi $. When $\mathbf{F}_{\text{BB}}^{\text{opt}}$ is decomposed into two sub-matrices $\mathbf{A}$ and $\mathbf{B}$, $\Phi ={{\Lambda }_{2}}-{{\Lambda }_{1}}$ and $\Phi >\text{0}$ which indicates that the number of floating-point operations is increased. Therefore, another two matrices $\mathbf{C}\in {{\mathbb{C}}^{{{N}_{\text{RF}}}\times m}}$ and $\mathbf{D}\in {{\mathbb{C}}^{m\times K}}$ have to be constructed to make $\Phi \le \text{0}$, i.e., the number of floating-point operations in the multiplication $\mathbf{CDs}$ is less than or equal to the number of floating-point operations in the multiplication ${{\mathbf{F}}_{\text{BB}}}\mathbf{s}$.

The SVD of $\mathbf{F}_{\text{BB}}^{\text{opt}}$ can be rewritten as
\begin{equation}
\begin{aligned}
  & \mathbf{F}_{\text{BB}}^{\text{opt}}=\mathbf{AB} \\
 & \ \ \ \ \ \ =\sum\limits_{i=1}^{K}{{{\mathbf{A}}^{\left( i \right)}}{{\mathbf{B}}_{\left( i \right)}}} \\
 & \ \ \ \ \ \ =\sum\limits_{i=1}^{K}{\left( {{\sigma }_{i}}{{\mathbf{u}}_{i}} \right)\mathbf{v}_{i}^{\text{H}}} \\
\end{aligned}
,
\end{equation}
where ${{\mathbf{A}}^{\left( i \right)}}$ is the $i\text{-th}$ column vector of $\mathbf{A}$ and ${{\mathbf{B}}_{\left( i \right)}}$ is the $i\text{-th}$ row vector of $\mathbf{B}$. Based on (10), $\mathbf{C}$ and $\mathbf{D}$ are constructed as
\begin{equation}
\mathbf{C}=\left[ {{\sigma }_{1}}{{\mathbf{u}}_{1}},{{\sigma }_{2}}{{\mathbf{u}}_{2}},\cdots ,{{\sigma }_{m}}{{\mathbf{u}}_{m}} \right]
\end{equation}
and
\begin{equation}
\mathbf{D}={{\left[ {{\mathbf{v}}_{1}},{{\mathbf{v}}_{2}},\cdots ,{{\mathbf{v}}_{m}} \right]}^{\text{H}}}
,
\end{equation}
where $m\in \left[ 1,K \right]$ is the number of column and row vectors in $\mathbf{C}$ and $\mathbf{D}$. The number of floating-point operations for the multiplication $\mathbf{CDs}$ is ${{\Lambda }_{3}}=8\left( {{N}_{\text{RF}}}m+mK \right)-2\left( {{N}_{\text{RF}}}+m \right)$. Considering the constraint $\Phi \le \text{0}$, i.e.,${{\Lambda }_{3}}\le {{\Lambda }_{1}}$, the value of $m$ satisfies

\begin{equation}
m\le \frac{{{N}_{\text{RF}}}K}{{{N}_{\text{RF}}}+K-\frac{1}{4}}.
\end{equation}
Based on (13), the value of $m$ is smaller than $K$, which indicates that the multiplication $\mathbf{CD}$ is the approximation of the multiplication $\mathbf{AB}$. Define the square of the Euclidean distance $\left\| \mathbf{AB}-\mathbf{CD} \right\|_{F}^{2}$ as the error $\left\| \mathbf{E} \right\|_{F}^{2}$, which is derived as
\begin{equation}
\begin{aligned}
  & \left\| \mathbf{E} \right\|_{F}^{2}=\left\| \mathbf{AB}-\mathbf{CD} \right\|_{F}^{2} \\
 & \ \ \ \ \ \ \ =\left\| \sum\limits_{i=m+1}^{K}{\left( {{\sigma }_{i}}{{\mathbf{u}}_{i}} \right)\mathbf{v}_{i}^{\text{H}}} \right\|_{F}^{2} \\
\end{aligned}
.
\end{equation}
Based on (14), the smaller the value of $m$ is, the larger the error will be. Furthermore, the performance deterioration in the throughput of mmWave massive MIMO systems depends on the value of $\left\| \mathbf{E} \right\|_{F}^{2}$. Considering that the value of $m$ is a positive integer, the maximum value of $m$ is given as
\begin{equation}
{{m}_{\max }}=\left\lfloor \frac{{{N}_{\text{RF}}}K}{{{N}_{\text{RF}}}+K-\frac{1}{4}} \right\rfloor,
\end{equation}
where $\left\lfloor x \right\rfloor $ is the floor function of a real number $x$, i.e., outputs the greatest integer which is less than or equal to $x$. When $m={{m}_{\max }}$, $\left\| \mathbf{E} \right\|_{F}^{2}$ is the smallest. On the contrary, $\left\| \mathbf{E} \right\|_{F}^{2}$ is the largest for $m=1$. An SVD-based decomposing (SVDDE) algorithm is proposed in the following to decompose the baseband precoding matrix into two sub-matrices. The proposed SVDDE algorithm is given as follows.

\begin{algorithm}
\caption{SVD-based decomposing (SVDDE)}
\begin{algorithmic}[1]
\REQUIRE {$\mathbf{F}_{\text{BB}}^{\text{opt}}$ and $\ensuremath{m}$}
\STATE Give the SVD of $\mathbf{F}_{\text{BB}}^{\text{opt}}$ and $\mathbf{F}_{\text{BB}}^{\text{opt}}=\mathbf{U\Sigma}\mathbf{V}^{\text{H}}$
\FOR {$\ensuremath{i=1,\cdots,m}$}
\STATE Select the $\ensuremath{i\text{-th}}$ column vectors of $\mathbf{u}_{i}$ and $\mathbf{v}_{i}$ which corresponding to the $i\text{-th}$ singular value $\ensuremath{\sigma_{i}}$
\STATE Set $\boldsymbol{\mathbf{C}}^{(i)}=\ensuremath{\sigma_{i}\mathbf{u}_{i}}$
and $\mathbf{D}_{(i)}=$$\mathbf{v}_{i}$
\ENDFOR
\STATE Generate matrices $\ensuremath{\mathbf{C}\text{=}\left[\sigma_{1}\mathbf{u}_{1},\sigma_{2}\mathbf{u}_{2},\cdots,\sigma_{m}\mathbf{u}_{m}\right]}$ and $\ensuremath{\mathbf{D}\text{=}\left[\mathbf{v}_{1},\mathbf{v}_{2},\cdots,\mathbf{v}_{m}\right]}$

\RETURN $\ensuremath{\mathbf{F}_{\mathbf{BB}}=\frac{\mathbf{CD}}{\left\Vert \mathbf{\mathbf{CD}}\right\Vert _{F}}}$

\end{algorithmic}
\end{algorithm}

\subsection{Computational Complexity Analysis}
The number of floating-point operations is considered as the computational complexity. One of the input in \textbf{Algorithm 1} is the near-optimal baseband precoding matrix $\mathbf{F}_{\text{BB}}^{\text{opt}}$, which is obtained by consuming ${{K}^{3}}+9{{N}_{\text{RF}}}{{K}^{2}}+3{{N}_{\text{RF}}}K$ floating-point operations. The SVD, in the first step of \textbf{Algorithm 1}, needs $4N_{\text{RF}}^{2}K+22{{K}^{3}}$ floating-point operations \cite{key-26}. Moreover, the number of floating-point operations in generating two matrices $\mathbf{C}$ and $\mathbf{D}$ is $4N_{\text{RF}}^{2}K+22{{K}^{3}}+2m{{N}_{\text{RF}}}$. Therefore, the computational complexity of \textbf{Algorithm 1} is $\Omega =9{{N}_{\text{RF}}}{{K}^{2}}+3{{N}_{\text{RF}}}K+4N_{\text{RF}}^{2}K+23{{K}^{3}}+2m{{N}_{\text{RF}}}$.

Based on the computational complexity of \textbf{Algorithm 1}, the computation power of BBUs is calculated and the total power consumption of the mmWave massive MIMO system is obtained. Therefore, the performance deterioration in the energy efficiency of mmWave massive MIMO systems can be analyzed.

\section{Simulation Results}
In this section, simulation results are provided to show the impact of decomposing $\mathbf{F}_{\text{BB}}^{\text{opt}}$ on the throughput and energy efficiency of mmWave massive MIMO systems with hybrid precoding structure. The carrier frequency is assumed to be 28 GHz. Moreover, there are 256 antennas and 60 RF chains in the mmWave massive MIMO system. Other default values of parameters in the mmWave massive MIMO system are listed in Table 1.

\begin{table}
\caption{Simulation parameters.}\label{tab1}
\begin{centering}
\begin{tabular}{|c|c|}
\hline
Parameters & Values\tabularnewline
\hline
Bandwidth $\ensuremath{W}$ & 1 GHz\tabularnewline

Transmission Power $P_{\textrm{T}}$ & 5 W\tabularnewline

Path loss exponent $\ensuremath{\alpha}$ & 4.6\tabularnewline

The number of multipath $\ensuremath{L}$ & 20\tabularnewline

Efficiency of PAs $\eta_{\textrm{PA}}$ & 38 \%\tabularnewline

Power of a phase shifter $P_{\textrm{PS}}$ & 12 mW\tabularnewline

Power of an RF chain $P_{\textrm{RF}}$ & 57 mW\tabularnewline
\hline
\end{tabular}
\par\end{centering}
\end{table}

\begin{figure}[htbp]
\centerline{\includegraphics[width=\textwidth]{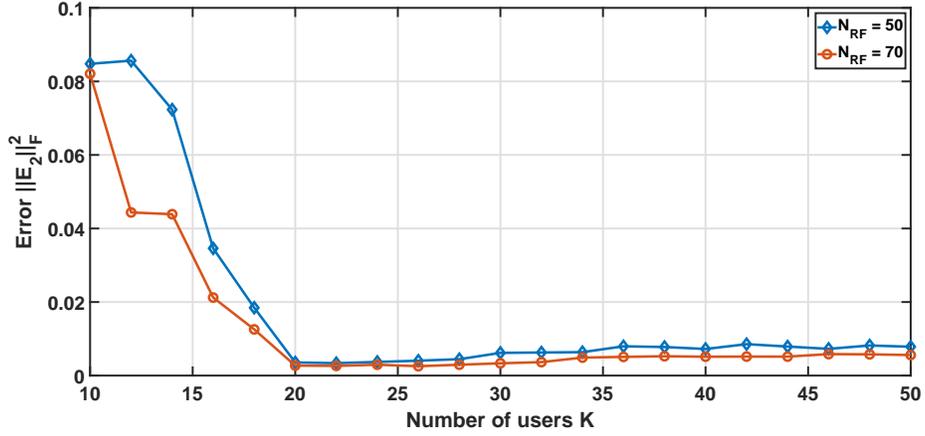}}
\caption{Error $\left\| {{\mathbf{E}}_{2}} \right\|_{F}^{2}$ with respect to the number of users.}\label{Fig3}
\label{fig}
\end{figure}

Fig. \ref{Fig3} shows the error $\left\| {{\mathbf{E}}_{2}} \right\|_{F}^{2}$ with respect to the number of users, considering different numbers of RF chains. The results in Fig. \ref{Fig3} indicate that the value of $\left\| {{\mathbf{E}}_{2}} \right\|_{F}^{2}$ first decreases with the number of users. However, there is an inflection point, i.e., $K=20$ for ${{N}_{\text{RF}}}=50$ and $K=22$ for ${{N}_{\text{RF}}}=70$, where the value of $\left\| {{\mathbf{E}}_{2}} \right\|_{F}^{2}$ starts increasing instead. Based on the results in Fig. \ref{Fig3}, there is an optimal number of users which can minimize the error. It is possible to use multi-layer neural networks to optimize the number of RF chains to minimize the error for different numbers of users.

\begin{figure}[htbp]
\centerline{\includegraphics[width=\textwidth]{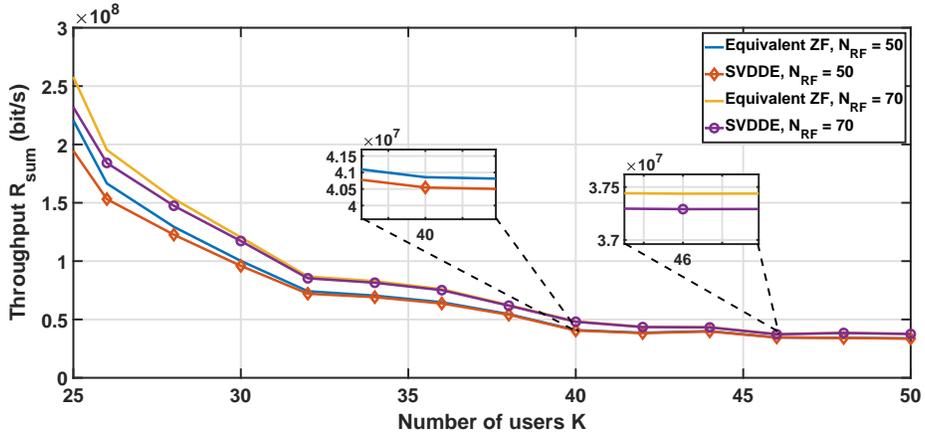}}
\caption{Throughput with respect to the number of users.}\label{Fig4}
\label{fig}
\end{figure}

In Fig. \ref{Fig4}, the throughput as a function of the number of users is illustrated when different numbers of RF chains is considered. Moreover, the equivalent ZF algorithm is simulated for performance comparison. The results in Fig. \ref{Fig4} indicates that the further decomposition of the baseband precoding matrix, based on the SVDDE algorithm, causes the performance deterioration in the throughput of mmWave massive MIMO systems with hybrid precoding structure. Based on the results in Fig. \ref{Fig4}, the gap between the equivalent ZF algorithm and the SVDDE algorithm in terms of the throughput is shrunk when the number of users is increased. Therefore, a slight performance deterioration in the throughput of multi-user mmWave massive MIMO systems is caused by further decomposing the baseband precoding matrix.

\begin{figure}[htbp]
\centerline{\includegraphics[width=\textwidth]{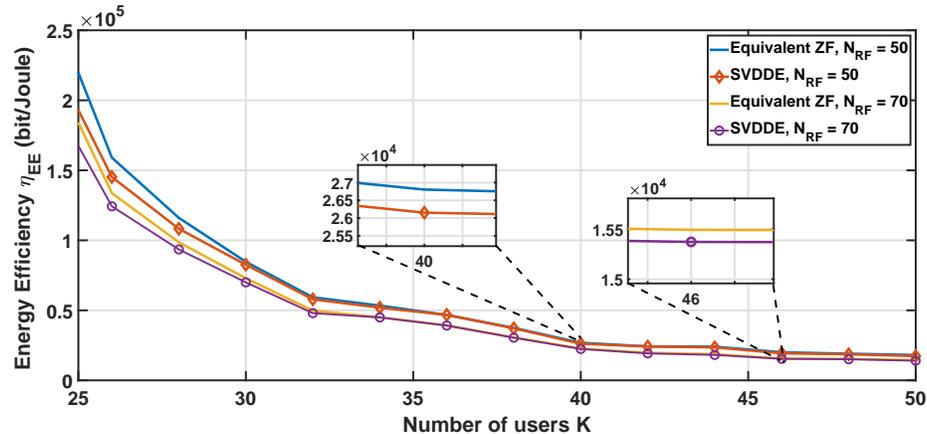}}
\caption{Energy efficiency with respect to the number of users.}\label{Fig5}
\label{fig}
\end{figure}

Fig. \ref{Fig5} illustrates the energy efficiency with respect to the number of users, considering different numbers of RF chains. Based on the results in Fig. \ref{Fig5}, the energy efficiency decreases with the number of users. One of two reasons for the decreasing of the energy efficiency is ascribed to the deterioration of the throughput in Fig. \ref{Fig4}. Based on (7), increasing the number of users grows the power consumption on BBUs, which is another reason for the decreasing of the energy efficiency. The gap between the equivalent ZF algorithm and the SVDDE algorithm in terms of the energy efficiency is narrowing with increasing the number of users in Fig. \ref{Fig5}. Therefore, a slight performance deterioration in the energy efficiency of multi-user mmWave massive MIMO systems is caused by further decomposing the baseband precoding matrix.

\section{Conclusions}
In this paper, an idea of further decomposing the baseband precoding matrix is proposed to map the hybrid precoding structure in mmWave massive MIMO systems to a multi-layer neural network. Moreover, an SVDDE algorithm is proposed to evaluate the feasibility of the proposed idea. Simulation results indicated that there is an optimal number of users which can minimize the performance deterioration. The existence of the optimal number of users shows the opportunity of applying multi-layer neural networks to design hybrid precoding. Moreover, simulation results also show that slight performance deterioration in the throughput and energy efficiency of mmWave massive MIMO systems is caused by further decomposing the baseband precoding matrix of hybrid precoding structure. Therefore, it is feasible to map the hybrid precoding structure of mmWave massive MIMO systems to a multi-layer neural network. For the future study, we will investigate the design of hybrid precoding based on software-hardware hybrid neural networks.

\end{document}